\documentclass[aps,prl,twocolumn,nofootinbib,amsfonts,floatfix]{revtex4}
\usepackage[colorlinks=true,linkcolor=blue,urlcolor=magenta, citecolor=blue]{hyperref}
\usepackage{amssymb,graphicx}
\usepackage{amsmath}
\usepackage{hyperref} 
\usepackage{upgreek}

\def\d{\partial}

\newcommand{\be}{\begin{equation}}
\newcommand{\ee}{\end{equation}}
\newcommand{\bea}{\begin{eqnarray}}
\newcommand{\eea}{\end{eqnarray}}
\newcommand{\bg}{\begin{gather}}
\newcommand{\eg}{\end{gather}}
\newcommand{\bseq}{\begin{subequations}}
\newcommand{\eseq}{\end{subequations}}

\begin{document}

\title{
 The QCD $\beta$-function On The String Worldsheet
 }
\author{ 
Sergei Dubovsky
}
\affiliation{
  Center for Cosmology and Particle Physics, Department of Physics,
      New York University,
      New York, NY, 10003, USA
      }

\begin{abstract}
We consider confining strings in pure gluodynamics and its extensions with adjoint (s)quarks.
We argue that there is a direct map between the set of bulk fields and the worldsheet degrees of freedom.
This suggests a close link between the worldsheet $S$-matrix and parton scattering amplitudes.  
We report an amusing relation between the Polchinski--Strominger amplitude responsible for the breakdown of integrability on
the string worldsheet and the Yang--Mills $\beta$-function
\[
b_0={D_{cr}-D_{ph}\over 6}\;.
\]
 Here $b_0=11/3$ is the one-loop $\beta$-function coefficient in the pure Yang--Mills theory, $D_{cr}=26$ is the critical dimension of bosonic strings and $D_{ph}=4$ is the dimensionality of the physical space-time we live in. A 
natural extension of this relation continues to hold in the presence of adjoint (s)quarks, connecting two of the most celebrated anomalies---the scale anomaly in quantum chromodynamics (QCD) and the Weyl anomaly in string theory.
\end{abstract}

\maketitle

Our description of strong interactions is embarrassingly incomplete without understanding of strings (flux tubes) responsible
for quark confinement. The stringy nature of the real world QCD manifests itself through the existence of Regge trajectories---families
of hadrons following a quadratic relation between the spin $J$ and the mass $M$,
\[
M^2\simeq J/\ell_s^2+const\;,
\]
where $1/\ell_s^2$ is the string tension. 
Critical string theory was born exactly 50 years ago \cite{Veneziano:1968yb} as an effort to explain this behavior.
Theoretical and lattice studies of confining strings are natural to perform in a more pristine environment obtained by eliminating dynamical quarks in the fundamental representation of the gauge group $SU(N_c)$. As a result, strings do not break  and one may study dynamics of an isolated infinitely long flux tube. In lattice simulations a long string state is created by the Polyakov loop operator \cite{Polyakov:1978vu}
\be
{\cal O}_P={\mbox \rm Tr} Pe^{i\oint A}\;,
\ee
wrapped around one of the spatial directions.

In the planar limit \cite{'tHooft:1973jz},  $N_c\to\infty$,  the worldsheet excitations decouple from bulk degrees of freedom and define a microscopic two-dimensional theory. 
Importantly, the worldsheet theory itself remains interacting even in the strict planar limit. Furthermore, there is mounting evidence that the worldsheet dynamics
is not described by a conventional local quantum field theory, but rather exhibits characteristic features of a gravitational theory \cite{Dubovsky:2018dlk}.

Much of the recent progress is triggered by identification of the worldsheet $S$-matrix as a primary fundamental observable \cite{Dubovsky:2012sh}.
This $S$-matrix is a natural theoretical target and at the same time has proven itself as an indispensable tool for the analysis of lattice data \cite{Dubovsky:2013gi,Dubovsky:2014fma}. 

Current lattice results \cite{Athenodorou:2010cs,Athenodorou:2011rx,Athenodorou:2013ioa,Athenodorou:2016kpd,Athenodorou:2016ebg,Athenodorou:2017cmw} (see
\cite{Teper:2009uf,Lucini:2012gg} for reviews) for both $D=4$ and $D=3$ gluodynamics can be summarized by the Axionic String Ansatz (ASA) \cite{Dubovsky:2015zey,Dubovsky:2016cog}. According to the ASA the only stable asymptotic degrees of freedom on the confining string  are massless Goldstone excitations $X^i$ ($i=1,\dots,D-2$) associated with spontaneous  breaking of translations in the presence of a long string. In addition, worldsheet scattering
at  $D=4$ exhibits a metastable resonance---the worldsheet axion \cite{Dubovsky:2013gi}. The axion is a pseudoscalar both w.r.t. the $O(2)$ group of rotations in the transverse plane, and w.r.t. the two-dimensional Poincar\'e symmetry $ISO(1,1)$ along the worldsheet. 

Both at $D=3$ and $D=4$ this is a matter content of an integrable theory enjoying the non-linearly realized target space Poincar\'e symmetry $ISO(1,D-1)$. The corresponding integrable phase shift coincides with the Dray--'t Hooft \cite{Dray:1984ha} gravitational shock wave phase shift
\be
\label{eils}
e^{2i\delta(s)}=e^{i\ell_s^2s/4}\;.
\ee
Exactly this phase shift describes integrable scattering on the worldsheet of critical (super)strings \cite{Dubovsky:2012wk}. It is also associated with a maximally chaotic behavior 
\cite{Maldacena:2015waa}.

  Both at $D=4$ and $D=3$ the integrability is not exact. At $D=4$ the absence of particle production requires the axion to be massless which is proven not to be the case by the lattice 
\cite{Athenodorou:2017cmw}. At $D=3$ one also finds clear deviations from integrability \cite{Dubovsky:2014fma,Dubovsky:2016cog} both in the  flux tube data
\cite{Athenodorou:2011rx,Athenodorou:2016kpd} and in the glueball spectra \cite{Athenodorou:2016ebg}.

On the other hand, the leading order coupling of the axion  determined from the lattice data \cite{Dubovsky:2013gi}  within theoretical and lattice uncertainties ({\it i.e.} at $\sim 10\%$ level) agrees with the value required for integrability 
\cite{Dubovsky:2015zey}. This suggests that the UV asymptotics of the worldsheet scattering may be governed by the shock wave phase shift (\ref{eils}). 

This proposal has a natural physical interpretation. The phase shift (\ref{eils}) corresponds to a time delay proportional to the collision energy, which may be taken as the most basic geometric property of a relativistic string. Given the underlying gauge theory is asymptotically free it is natural to assume that the high energy limit of the worldsheet scattering is largely determined by these geometric considerations.

To test the ASA further one needs to understand how asymptotic freedom of the bulk theory translates into the properties of 
 the high energy worldsheet scattering. As phrased so far the worldsheet dynamics appears to be rather disconnected from perturbative QCD.
 
\begin{figure}[t!]
  \begin{center}
        \includegraphics[width=7cm]{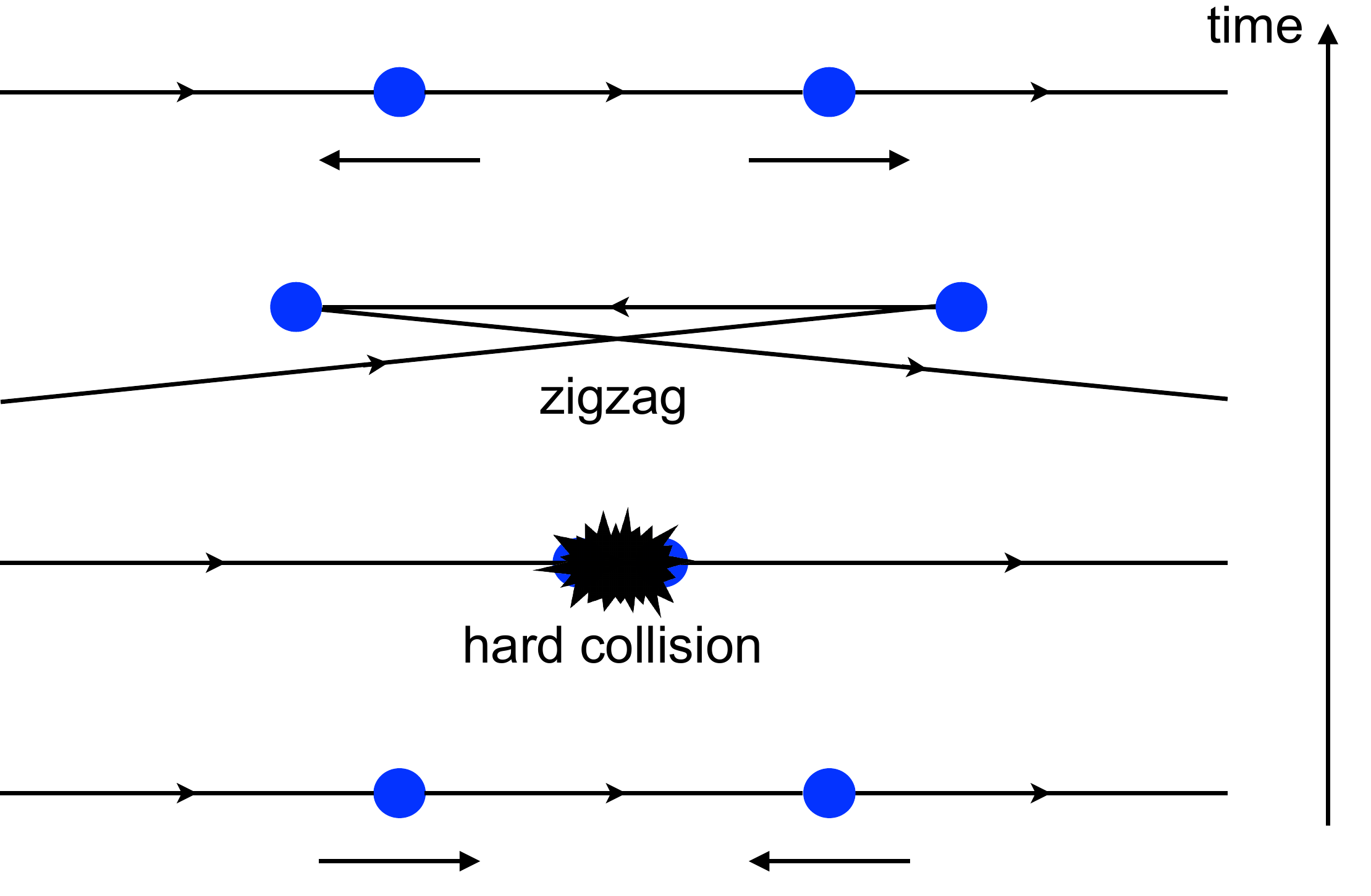} 
           \caption{High energy worldsheet scattering at $D=2$ proceeds through a hard collision followed by a prolonged zigzag phase.}
        \label{fig:zigzag}
    \end{center}
\end{figure}

 Recently a progress in this direction was achieved via the analysis of the $D=2$ case \cite{Dubovsky:2018dlk}. A pure Yang--Mills theory at $D=2$ is topological and exactly solvable even at finite $N_c$ \cite{Migdal:1975zg,Kazakov:1980zi,Kazakov:1980zj,Rusakov:1990rs,Witten:1991we,Gross:1992tu}. To introduce local dynamics one considers, following \cite{Dalley:1992yy,Bhanot:1993xp,Kutasov:1993gq}, a version of the model with additional massive adjoint (s)quarks.
At heavy (s)quark masses, $m\ell_s \gg 1$, the model can be treated perturbatively. The worldsheet theory arises as a subsector in a discrete $\theta$-vacuum \cite{Witten:1978ka}. Each adjoint (s)quark field $\psi$ maps into a color singlet excitation
on the worldsheet. The color flux of (s)quarks can be thought to be screened by infinitely heavy fundamental charges at spatial infinity, which  produce a flux tube.
Multiparticle states on the worldsheet are created by single trace operators of the form
\be
\label{psis}
{\cal O}_n={\mbox \rm Tr} Pe^{i\oint A}\psi_1\dots\psi_n\;,
\ee
where $n$ is the number of particles.
High energy scattering on the worldsheet is indeed dominated by  time delays proportional to the collision energy. The scattering proceeds through the formation of zigzag configurations, see Fig.~\ref{fig:zigzag}, which are responsible for the geometric time delay. When an elusive gravitational description of the worldsheet  dynamics is achieved zigzags are expected to map into black holes.

Focusing on the worldsheet dynamics brings in an advantage that the worldsheet theory always lives in two dimensions independently of the dimensionality $D$ of an 
underlying gauge theory. This makes it straightforward to uplift the knowledge gained in the analytically tractable $D=2$ case into higher dimensions. Of course, the presence of massless gluons precludes a direct perturbative analysis at $D=3,4$ in certain regimes. However, it is not unreasonable to expect the major qualitative features present in the perturbative regime to survive also at strong coupling.

In particular, analogously to (\ref{psis}), in $D>2$ pure glue theories  the wordsheet excitations are created by inserting the gluon field strength inside the Polyakov loop.  Let us consider a long confining string stretched along $z$ direction in the $D=4$ case. Then one expects to find one-particle excitations corresponding to operators
\be
\label{Os}
{\cal O}_i={\mbox \rm Tr} Pe^{i\oint_z A}F_{zi}\;,\;\; {\cal O}_{a}={\mbox \rm Tr} Pe^{i\oint_z A}F_{ij
}\;,
\ee
where $i=x,y$ label transverse spatial directions. ${\cal O}_i$ operators match quantum numbers of the Goldstone modes and are guaranteed to produce massless worldsheet excitations.  The ${\cal O}_a$ operator matches the quantum numbers of the worldsheet 
axion. The corresponding excitation is not protected and expected to acquire a mass and to be unstable. In the lattice description operators (\ref{Os}) are obtained by inserting into the Polyakov loop either a plaquette along one of the longitudinal directions (${\cal O}_i$'s), or in the transverse plane (${\cal O}_a$). At $D=3$ one is left with a single longitudinal plaquette.

This largely demystifies the ASA---it reduces to the statement that the wordsheet theory has the minimal excitation spectrum compatible with the bulk matter content.
This also provides a dual view of the Goldstone modes. Low energy Goldstone modes, as well as their coherent multiparticle  excitations, are most appropriately described by geometric deformations of the string worldsheet. On the other hand, hard one-particle Goldstone excitations can be thought of as gluons. The geometric phase shift (\ref{eils}) arises in both descriptions, even though the detailed underlying pictures are a bit different. In the Goldstone language the time delay corresponding to (\ref{eils}) arises 
as a consequence of the linear relation between the energy of a string segment and its proper length $L$  \cite{Dubovsky:2012wk},
\[
L=\ell_s^2E\;.
\]
In the  gluon description it comes about in the same way as in the $D=2$ case---hard gluons overshoot each other, proceed through the zigzag stage, and eventually get turned back by a long string stretched between them. This again results in a time delay proportional to the collision energy.  Both mechanisms are  semiclassical in  nature, however, they are quite different. In particular, the first one corresponds to a pure transmission, while the latter is a total reflection. This difference is not observable for identical particles. However, if we replace one of the gluons with an adjoint (s)quark the difference becomes physically detectable.

To see yet another distinction consider, for simplicity, the $D=3$ case when the worldsheet carries a single stable  (and massless) excitation, and write a dispersion relation for the corresponding two-particle 
$S$-matrix $S_2(s)$,
\be
\label{disp}
\oint ds {S'_2(s)\over s S_2(s)}=2\pi i \sum_{zeros} {1\over s}\;.
\ee
\begin{figure}[t!]
  \begin{center}
        \includegraphics[width=7cm]{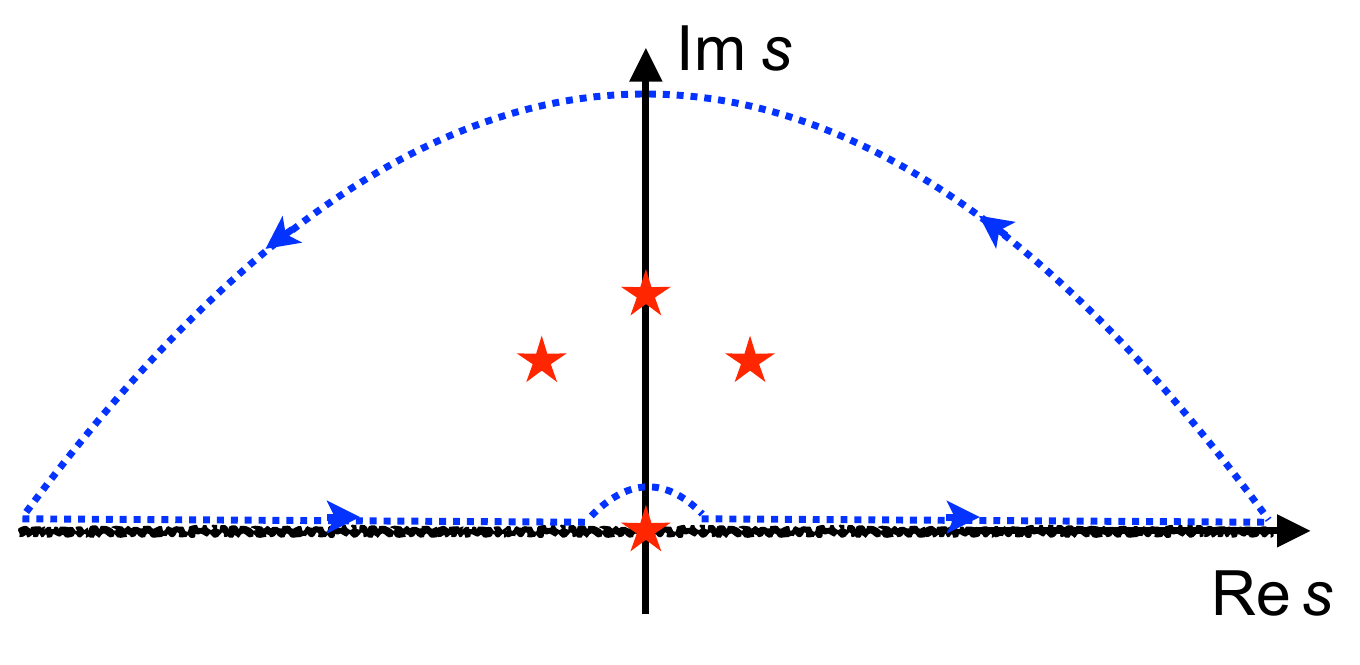} 
           \caption{An integration contour for the dispersion relation (\ref{disp}).  }
        \label{fig:zeros}
    \end{center}
\end{figure}
Here the integration contour goes around the upper half-plane of the Mandelstam variable $s$, and the sum in the r.h.s. is over zeros of $S_2$ there, see Fig.~\ref{fig:zeros}. Very similar dispersion relations appear in the derivation of the superluminality bound \cite{Adams:2006sv}, in the proof of the $a$-theorem \cite{Komargodski:2011vj} and in the recent work on the $S$-matrix bootstrap \cite{Paulos:2016but,Doroud:2018szp}.
The integral in (\ref{disp}) receives contributions from the pole at $s=0$ where 
\[
S_2({s\to 0})\approx 1+i \ell_{IR}^2 s/4+\dots\;,
\]
 from the cut along the real axis and from the semicircle at infinity, where
 \[
 S_2({s\to \infty})\sim e^{i\ell_{UV}^2s/4}\;.
 \]
Altogether, (\ref{disp}) translates into the following positivity bound,
\be
\label{positive}
\ell_{IR}^2-\ell_{UV}^2=-{4\over \pi}\int_0^\infty{\log{ |S_2|^2}\over s^2}+8i\sum_{zeros}{1\over s}\geq0\;.
\ee
The inequality in (\ref{positive}) follows from unitarity (implying that the integral term is non-negative) and from crossing symmetry, which ensures that each zero at $s_0$ is either purely imaginary or accompanied by another one at $-s_0^*$ (implying that the sum term is non-negative).

We see that the time delay due to hard zigzag scattering (controlled by $\ell_{UV}^2$) is always shorter than the time delay  characterizing scattering of soft semiclassical modes of the same total energy (controlled by $\ell_{IR}^2$). 
At first sight this mismatch is inconsistent with the simple geometric picture of scattering advocated above, where the time delay is always controlled by the tension of a long string ({\it i.e.}, by $\ell_{IR}^2$).

However, the discrepancy arising due to the integral term in (\ref{positive}) has a natural physical interpretation. The integral term  is related to particle production, which may force colliding gluons to turn around earlier than in a purely elastic regime. This may lead to a faster termination of the
zigzag stage. 

Interestingly, the string length $\ell_s^2$ in the $D=3$ Yang--Mills, determined  by fitting the
slope of the leading Regge trajectory of low lying glueballs, is significantly (by a factor of $\sim 1.27$) smaller than the value of $\ell_s^2$ measured from the 
ground state energy of a long flux tube \cite{Dubovsky:2016cog}. However, the latter corresponds to $\ell_{IR}^2$, while the former is more naturally associated with $\ell_{UV}^2$, so the bound (\ref{positive})  suggests a natural resolution of this puzzle.

 It will be interesting to see what this implies for the spectrum of particle produced in the worldsheet scattering. It should be possible to estimate its properties given that the zigzag stage is characterized by a long period of  constant acceleration, suggesting the possibility of a quasithermal spectrum. This is another clear call for a gravitational reformulation of the theory. 
 
 Let us  point out yet another geometric source of soft particle production, which should also be possible to account for. At $D>2$ one does not expect the hard collision to be exactly collinear. There always will be a (small) scattering angle. As a result the zigzag is not precisely aligned with the string, which translates in a certain emission spectrum
 of soft Goldstones.

On the other hand, it appears impossible to accommodate the  zeros' contribution in (\ref{positive}) into a geometric description of scattering. 
In fact, as proven in \cite{Dubovsky:2015zey}, zeros are absent in the integrable case, leaving the shock wave $S$-matrix as the only option for an integrable $D=3$ $S$-matrix compatible with the non-linearly realized Poincar\'e symmetry. It will be interesting to see whether zeros may be excluded  from first principles also in a non-integrable case. If so, this will provide a sharp version of the $D=3$ ASA, which is actually well supported by the glueball spectroscopy \cite{Dubovsky:2016cog}.

Note that the $D=3$ $k$-string lattice data \cite{Athenodorou:2013ioa} does show the presence of massive resonances \cite{Dubovsky:2014fma}. However, these should disappear in the planar limit, when the worldsheet theory becomes
UV complete.  In the planar limit a $k$-string reduces simply to $k$ decoupled copies of a fundamental string (assuming $k$ is kept fixed; it is unclear whether $N_c\to\infty$ limit with fixed $k/N_c$ gives rise to a microscopic 2d theory).

We see that a copious production of soft particles is likely to play an important role in understanding the worldsheet scattering. This is especially natural in view of the following reformulation of the gluon/Goldstone duality\footnote{We thank Riccardo Rattazzi for suggesting this very instructive viewpoint.}. It is instructive to think of a long string as a very special highly symmetric hadronic state. Hard colliding gluons (and adjoint (s)quarks, if present) are nothing but  valent partons of this hadron. The ground state of an infinitely long string corresponds to a special hadron with no valent dynamical partons at all (apart from the external charges at infinity). Given the crucial importance of soft and collinear gluon emission in hadron physics it is hardly surprising that soft inelastic processes play an important role on the worldsheet. 

\begin{figure}[t!]
  \begin{center}
        \includegraphics[width=7cm]{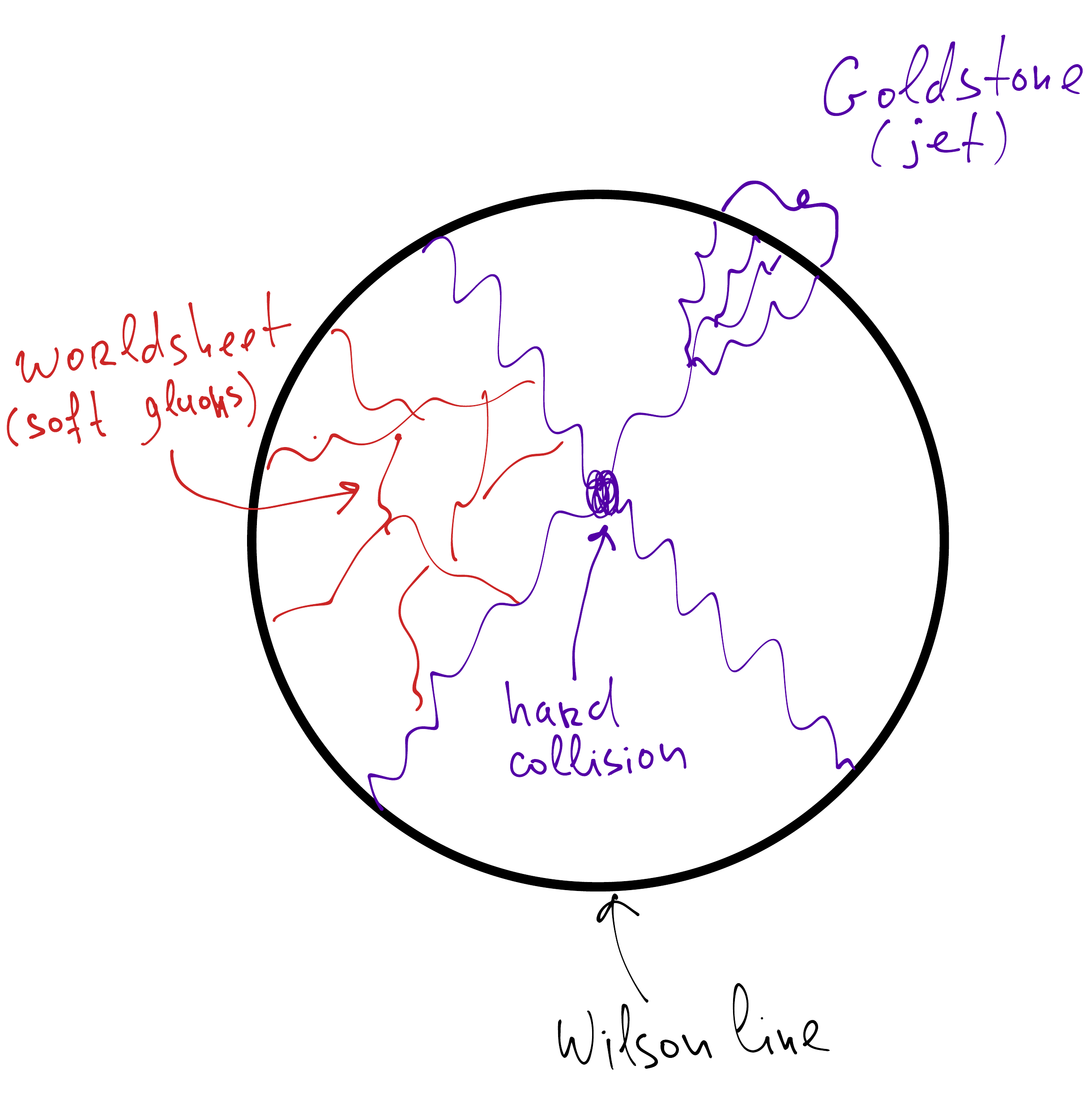} 
           \caption{A very schematic drawing of the Wilson line as an IR regulator. We fail to properly draw this using double line notations. Also this is a Feynman diagram rather than a space-time picture, so that the zigzag region is not manifest, even though it is physically present.}
        \label{fig:softcol}
    \end{center}
\end{figure}

Given that the Goldstone amplitudes are IR finite, a yet another way to think about the present setup is that in the planar limit a Wilson line at infinity, associated with a pair of external charges, provides a very special IR regulator enforcing strictly collinear kinematics.
In this language massless Goldstones correspond to jets of hard gluons dressed by collinear radiation, while soft gluons form the string worldsheet, see Fig.~\ref{fig:softcol}.  

All of these viewpoints strongly suggest that the worldsheet scattering is related to perturbative QCD (including, in particular, gluon scattering amplitudes and soft and colliniear  splitting functions) in a very direct way. We feel that a detailed understanding of this relation is the next natural step in solving the riddle of confining strings.

As a first step in this direction let us revisit two-particle scattering on the worldsheet with an eye on a possible connection to perturbative QCD. The key characteristic
feature of the tree level $2\to 2$ scattering in the $D=4$ Nambu--Goto theory is the absence of annihilations and reflections---the tree level $2\to 2$ $S$-matrix describes pure transmission \cite{Dubovsky:2012sh}. It is natural to reformulate this property in the helicity basis. Let us introduce complex combinations of the Goldstone fields
\[
X=X^x+i X^y\;,\;\; \bar X=X^x-i X^y\;,
\]
where as before we are considering a long string stretched in $z$ direction. Then $\d_+X$ and $\d_-\bar X$ correspond to helicity plus string excitations, and
$\d_+\bar X$ and $\d_- X$ to helicity minus (here $\d_\pm=\d_t\pm\d_z$). As a consequence of pure transmission the  $\d_+X\d_-\bar X\to \d_-X\d_+\bar X$ amplitude vanishes. Interestingly, the tree level  $2\to 2$ gluon amplitude also exhibits the same property (see, e.g.,  \cite{Dixon:2013uaa} for a review), 
\[
A^{tree}_4(+,+,+,+)=0\;.
\] 

At the moment it is hard to tell whether this similarity is coincidental or not. Clearly, the two calculations have very different regimes of applicability. The Goldstone calculation applies at  the leading order in derivative expansion, while the gluon result is a tree level approximation applicable at high energies when the gauge theory description is weakly coupled. Note that multiparticle tree-level Nambu--Goto amplitudes  are integrable ({\it i.e.}, there is no particle production). It will be interesting to understand what is the counterpart of this integrability in the multigluon scattering, if any.

The Nambu--Goto integrability is broken at the one-loop order by a universal rational term  \cite{Dubovsky:2012sh}. This term is closely related to the Weyl anomaly of non-critical strings \cite{Polyakov:1981rd} and was first derived by Polchinski and Strominger (PS) \cite{Polchinski:1991ax}, even though at the time it was not recognized as a contribution to the scattering amplitude (a modern exposition of the PS formalism is presented in \cite{Hellerman:2014cba}, and its precise relation to the worldsheet scattering is explained in \cite{Dubovsky:2016cog}). At the level of two-particle scattering this term translates into the following annihilation amplitude (we use the same normalization as in 
\cite{Dubovsky:2014fma}),
\be
\label{ann}
A_{ann}= {26-D\over 24\pi} {\ell_s^4 s^2\over 16}\;.
\ee
Given the present context it is impossible to ignore that at $D=4$ the prefactor in (\ref{ann}) coincides with the gluonic contribution into the QCD $\beta$-function
\cite{Gross:1973id,Politzer:1973fx},
\be
\label{beta}
\beta(\alpha_s)=-{22-n_{sc}-4n_f\over 24\pi}C_A\alpha_s^2\;,
\ee
where we used the PDG conventions \cite{Patrignani:2016xqp} and included also a contribution from $n_{sc}$ Hermitian adjoint squarks and $n_f$ Weyl adjoint quarks.
As a zeroth order check that this coincidence is not an obvious numerology let us see whether massless adjoint (s)quarks  affect the PS amplitude in the same way as the $\beta$-function.

Following the mapping (\ref{psis}) a Hermitian adjoint squark translates into an additional real scalar field $\phi$ on the worldsheet. Its leading order interactions with the Goldstones are
\be
\label{Sphi}
S_{\phi}=-{1\over 2}\int\sqrt{-h}h^{\alpha\beta}\d_\alpha\phi\d_\beta\phi\;,
\ee
where $h_{\alpha\beta}$ is the induced metric. The corresponding $XX\phi\phi$ vertices are the same as one would get from expanding the Nambu--Goto action in $D+1$ dimensions. Hence, the calculation of the one-loop scattering of Goldstones proceeds exactly as in  \cite{Dubovsky:2012sh}. In this calculation $\phi$ acts now as an additional 
spatial dimension, shifting  $D$ into $D+1$ in (\ref{ann}). This agrees with (\ref{beta}).

In general, fermions can be incorporated on the worldsheet  following the coset construction \cite{Mohsen:2016lch}. However, given that in the case at hand they come in complete multiplets of the target space Poincar\'e group, one can take a shortcut and immediately write the corresponding leading order action as
\be
\label{Sfer}
S_\Psi=\int\sqrt{-h}\bar\Psi\Gamma_\mu\d_\alpha X^\mu\d_\beta\Psi h^{\alpha\beta}\;.
\ee
The fastest way to see that bulk fermions affect (\ref{ann}) in the same way as (\ref{beta}) is to note that for the $N=4$ superconformal Yang--Mills (so that $\beta=0$) the mapping (\ref{psis}) gives rise to the same matter content on the worldsheet as for the critical type IIB superstrings,
where the PS interaction vanishes. This fixes also the fermionic contribution into (\ref{ann}) to be the same as in (\ref{beta}). The same conclusion follows also from a direct one-loop calculation presented in \cite{Mohsen:2016lch}.

Given that the sign of the $\beta$-function is related to asymptotic freedom it will be interesting to see whether the sign of (\ref{ann}) can also be constrained from a dispersion relation similar to (\ref{positive}).

Note that at $D=3$ where the gauge theory does not exhibit any logarithmic running, the PS amplitude is also identically zero for kinematical reasons.

If not a random coincidence, what is the possible physical origin for the agreement between (\ref{ann}) and (\ref{beta})?
At first sight, this relation has suggestive similarities with the anomaly matching. The one-loop $\beta$-function controls the leading logarithmic violation of the scale invariance at high energies, and the PS interaction is related to the Weyl anomaly in the low energy effective theory on the worldsheet. Note, however,
that the combination $(n_s+4n_f)$ is not equal to the central charge of the low energy theory on the worldsheet. The PS amplitude would be proportional to the central charge if the worldsheet fermions were singlets under the $O(D-2)$ group of transverse rotations (as it happens for the nonsupersymmetric sector of 
heterotic  strings in the fermionic description). Contribution of non-singlet fermions to the PS interaction is different, which in this language explains, for instance, why the critical central charge for superstrings is $c=15$ rather than $c=26$.

An even more important point is that conformal symmetry is broken by the RG flow, so at least as far as this symmetry is concerned, 
one does not expect to find  anomaly matching, but rather an inequality at best, with $a$- and $c$-theorems \cite{Zamolodchikov:1986gt,Cardy:1988cwa,Komargodski:2011vj} serving as the celebrated  examples. Somewhat related to this, even though massless bulk and worldsheet (s)quarks affect the one-loop $\beta$-function and  the PS amplitude in the same way, generically (say, without supersymmetry) they will not stay massless on the worldsheet, making it hard to make a sharp non-perturbative statement.

For these reasons, it  appears more natural to look for the explanation of the agreement between (\ref{ann}) and (\ref{beta}) in the perturbative dynamics directly related to the asymptotic freedom, which would  actually be in line with the earlier logic which lead us here.
In this regard, note that
the one-loop  $\beta$-function appears as a prefactor in front of $\delta(1-x)$ contribution into the leading order Altarelli--Parisi gluon splitting function (see, e.g., \cite{Ellis:1991qj}).
Given that the worldsheet scattering is closely related to collinear physics, this looks as a natural dynamical route for the QCD $\beta$-function to propagate into the PS amplitude.

As phrased inititally the PS/$\beta$-function equality appears as the agreement between an UV quantity (one-loop $\beta$-function) and an IR quantity (the PS amplitude). However, in view of this discussion it is probably more appropriate to consider it as the agreement between two UV quantities. This provides the sharpest formulation of the equality. Namely, it states that the PS amplitude calculated in the worldsheet theory with matter content determined by the mapping (\ref{psis}) and all masses set to zero (turning this into a UV statement) is equal to the one-loop $\beta$-function coefficient.

Note that the  coincidence between the one-loop $\beta$-function coefficient and  the worldsheet Weyl anomaly has been observed previously in 
\cite{Fradkin:1982kf,Metsaev:1987ju,DiVecchia:1995in,Rey:1997hj}. The logic of these works is very different though. There the connection arises by embedding the gauge theory into a string theory and taking the $\alpha'\to 0$ limit.
Here, instead, it is a statement about two calculations done intrinsically within QCD.

To conclude, there are multiple reasons to expect a close
connection between the worldsheet scattering and perturbative QCD. 
 Understanding the details of this relation looks as a natural continuation of the confining string saga. We anticipate it to be as exciting as the path which brought us to the present point.

{\it Acknowledgements.} We are especially  grateful to Riccardo Rattazzi for numerous lively and stimulating discussions. We thank Raphael Flauger, Victor Gorbenko and Guzm\'an Hern\'andez-Chifflet for
long time collaboration on closely related topics. We also benefited from conversations with Giga Gabadadze, Volodya Kazakov, Ali Mohsen, Gabriele Veneziano and Xi Yin. 
This work is supported in part by the NSF CAREER award PHY-1352119.

\bibliographystyle{h-physrev3}
\bibliography{dlrrefs}
\end{document}